# How much does surface polymorphism influence the work function of organic/metal interfaces?

Andreas Jeindl[1], Lukas Hörmann[1], and Oliver T. Hofmann[1]*

[1] Institute of Solid State Physics, NAWI Graz, Graz University of Technology, Petersgasse 16, 8010 Graz, Austria

* Corresponding Author:
Oliver Hofmann - Institute of Solid State Physics, NAWI Graz, Graz University of Technology, Petersgasse 16, 8010 Graz, Austria; orcid.org/0000-0002-2120-3259; Phone: +43 316873 8964; Email: o.hofmann@tugraz.at

**Keywords**: *density functional theory, structure prediction, machine learning, kinetic trapping, ab-initio thermodynamics, temperature dependence*

**ABSTRACT**

Molecules adsorbing on metal surfaces form a variety of different surface polymorphs. How strongly this polymorphism affects interface properties is a priori unknown. In this work we investigate how strongly the surface polymorphism influences the interface work functions for various metal/organic interfaces. To evaluate the whole bandwidth of possible polymorphs, we perform full theoretical structure search, probing millions of polymorph candidates. All of these candidates might be observed in reality, either by kinetic trapping or by thermodynamic occupation. Employing first-principles calculations and machine learning we predict and analyze the work function changes for those millions of candidates for three physically distinct model systems: the weakly interacting naphthalene on Cu(111), the strongly interacting anthraquinone on Ag(111), and tetracyanoethylene, which undergoes a re-orientation from lying to standing polymorphs on the Cu(111) surface. These thorough investigations indicate that kinetic trapping of flat lying molecules can lead to work function differences of a few hundred meV. If the molecules also reorientate, this can increase to a change of several eV. We further show that the spread in work function decreases when working in thermodynamic equilibrium, but thermally occupied phases still lead to an intrinsic uncertainty at elevated temperatures.

**TOC Graphic**

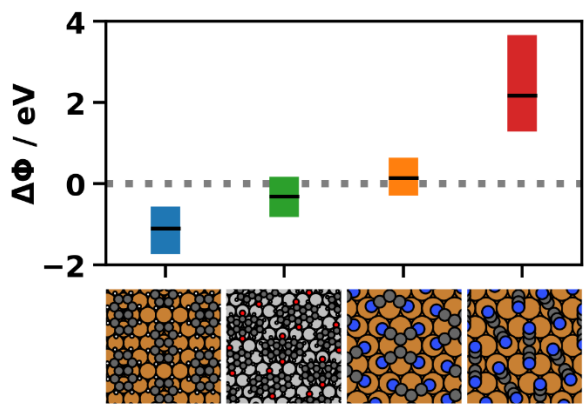

# 1. Introduction

A big contribution to the device performance of organic electronic devices is given by the charge injection barriers from the inorganic contact into the organic active material.[1–3] In a simple picture, these barriers are given by the electronic level alignment, i.e. the offset between the transport levels of the active material and the work function of the electrode, $\phi$.[3] An efficient way to improve the efficiency of organic devices, therefore, is to modify $\phi$. For metal electrodes, this can be done via the application of so-called charge-injection layers,[1,4–6] e.g. in the form of organic molecules that introduce a potential jump $\Delta\Phi$ above the metal surface.

A major challenge of this approach, however, is that organic molecules show extensive polymorphism, with different polymorphs often exhibiting strongly different properties. A prime example of this behavior is pentacene, for which 5 different phases (on multiple surfaces) with strongly differing charge-carrier mobilities are known.[7–11] Which structure forms in an actual experiment depends heavily on the processing conditions. Even when very similar conditions are employed, the outcome can differ considerably: Here, a classical example is PTCDA on Ag(111), for which different studies reported work functions differing by approx. 200 meV [12,13] (although it has not yet been confirmed whether this is due to polymorphism or stems from other sources). The variance of properties may be because physical vapor deposition frequently leads to kinetically trapped phases[14] or, even if this is not the case, because the energy differences between different polymorphs are so small that multiple structures may be thermally occupied.[15] In this work, we assess how large this effect can become, i.e., how much the $\Delta\Phi$s of different polymorphs differ from each other, and whether systems that display a certain kind of interaction with the surface are more or less susceptible to it.

For a broad overview we investigate three complimentary classes of molecules on metal surfaces: i) molecules that interact only weakly (e.g. via van der Waals forces) with each other and with the surface, ii) molecules which undergo a charge transfer reaction on the surface, leading to an electronic coupling between molecule and substrate, and iii) molecules which undergo a large structural change when changing the deposition conditions. Combining recent advances in machine learning with state-of-the-art density functional theory (DFT) calculations, we predict both the energies and the $\Delta\Phi$s of millions of possible polymorphs candidates. This allows us to explore how strongly the work function of a given interface may differ between different experiments if the deposition of the molecules is performed either under conditions that favor kinetic trapping or when the growth occurs in thermodynamic equilibrium.

# 2. Methods

As basis for our machine learning approach, we use the adsorption energies and adsorption induced work-function modifications of several hundred polymorph candidates for each of the three interfaces (251 for naphthalene/Cu, 245 for A2O/Ag, 319 for TCNE/Cu). The polymorphs candidates were created using the SAMPLE[16] approach (a structure search package specifically designed for inorganic/organic interfaces), allowing for up to 6 non-equivalent molecules per unit cells, without constraints on the shape of the unit cell. The calculations for these candidates, which are also available online, [17–21] were consistently performed as periodic slab-type calculations, employing the exchange-correlation functional PBE[22] in combination with the TS$^{surf}$ correction[23,24] to account for long-range dispersion interactions. Within this work, we define the adsorption energy as $E_{ads} = E_{sys} - E_{sub} - E_{mol}$ where $E_{sys}$ is the energy of the combined system, $E_{mol}$ the energy of a molecule in the gas phase, and $E_{sub}$ the energy of the pristine metal slab. Negative values of $E_{ads}$ denote energy gain upon adsorption. The adsorption-induced work function change $\Delta\Phi$ is defined as the difference between the work function of the interface (metal substrate with a monolayer of adsorbates) and the pristine metal substrate.

To exhaustively predict the properties of millions of different polymorph candidates (with different numbers of molecules in different unit cells), we employ the SAMPLE[16] approach which utilizes the following model:

$$X = \sum_{geoms} N_g U_g + \sum_{pairs} N_p V_p \qquad (1)$$

The model consists of a linear combination of molecule-substrate ($U_g$) and molecule-molecule interactions ($V_p$), each multiplied with the number of occurrence for this specific interaction ($N_g$ and $N_p$) in the polymorph. Using the above-mentioned DFT calculations, we then obtain all interaction parameters of equation 1 via Bayesian linear regression. This approach has been previously successfully employed to predict the adsorption energies of polymorphs candidates with the same accuracy as the underlying electronic structure method. [25–27] Here, we extent the application to adsorption-induced work function changes ($\Delta\phi$).

For this, we need to apply one further step: The work function is an intensive property which depends on the coverage of the different polymorph candidates. As regression directly based on $\Delta\phi$ would violate its low-coverage limit (for very low coverages $\Delta\phi$ would converge towards the $U_g$ instead of zero), we instead train our machine-learning model and predict the dipole per adsorbed molecule. The dipole per unit cell directly relates to $\Delta\phi$ via the Helmholtz solution to the Poisson equation (equation (2))

$$\Delta\phi = -\frac{\mu}{\epsilon_0 A} \qquad (2)$$

Hereby $\epsilon_0$ is the dielectric constant, $A$ the area of the unit cell and $\mu$ the resulting dipole of the structure. This approach allows us to calculate $\Delta\phi$ with DFT, convert it to dipoles via equation (2), train and predict the dipoles, and then convert them back via the same mathematical relationship to obtain the $\Delta\phi$ for all polymorph candidates of all systems.

The SAMPLE approach relies on a set of hyper parameters. For the energy prediction we utilized the hyper parameters of the original publications. [25–27] To predict the dipoles, of course, different parameters need to be employed. The detailed values used are given in Table S1 and Table S2 of the Supporting Information. The uncertainties of the work functions evaluated via leave one out cross validation (LOOCV) are: 5 meV for naphthalene, 8 meV for A2O, 10 meV for TCNE flat, and 80 meV for TCNE standing (however, the range for TCNE standing is also by a factor of 5-10 larger than for the other systems).

For the calculation of the Gibbs free energy of adsorption γ we employed ab initio thermodynamics.[28] In doing this, we neglect contributions of the mechanical work, the configuration entropy and the vibration enthalpy, as is commonly done in literature[28,29]. This approach leads to

$$\gamma_i = (E_i^{ads} - \mu(T, p))/A_i \qquad (2)$$

Here, $\mu(T, p)$ is the chemical potential of the molecules in gas phase at a specific temperature and pressure. $A_i$ represents the area per molecule for each separate polymorph candidate. Within this approximation, the temperature dependence is introduced via the chemical potential of the molecules in gas phase. Note that we define $E_{ads}$ and $\mu(T)$ both in terms of energy per molecule, which makes those values independent of the number of molecules per unit cell. We acknowledge that the vibrational entropy can play a role when directly comparing stabilities of systems with strongly varying molecule-surface interactions. For our systems, this would only be relevant when directly comparing the lying and standing TCNE configurations, which has been done elsewhere.[25] For further details of our ab-initio thermodynamics approach see [27].

## 3. Results and Discussion

For a broad overview of the influence of the possible polymorphism on the work functions of organic/metal interfaces, we consider three different material classes (illustrated in Figure 1).

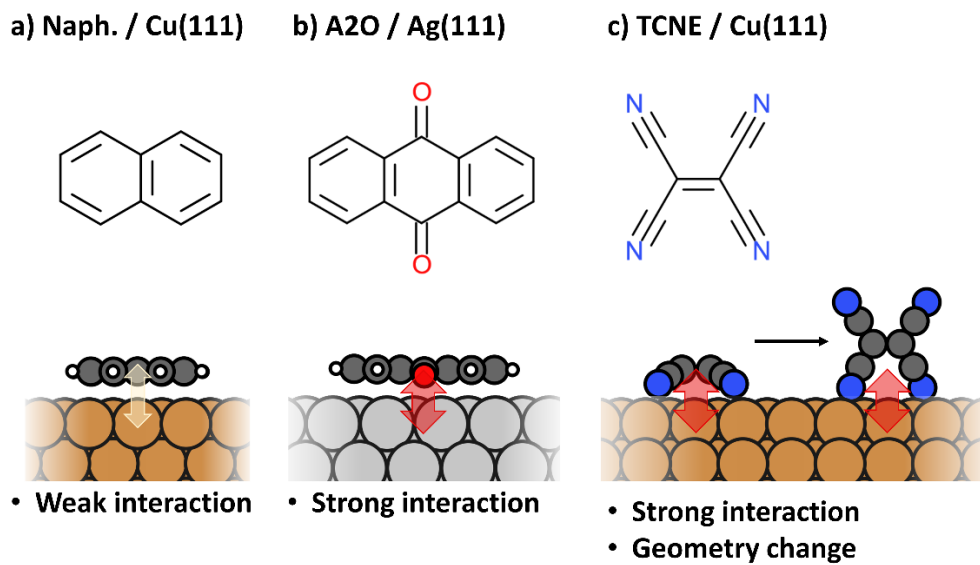

**Figure 1:** Systems used for this study. a) naphthalene on a Cu(111) surface, b) 2,7-anthraquinone (A2O) on a Ag(111) surface and c-d) tetracyanoethylene (TCNE) lying and standing on a Cu(111) surface.

Weakly interacting: Our discussion starts with the behavior of naphthalene on Cu(111), a system that is only weakly interacting but is well known to exhibit pronounced polymorphism.[16,30,31]

Strongly interacting: Second, we discuss the behavior of acenequinones on Ag(111). These molecules exhibit a complex charge transfer reaction with the substrate, consisting of the formation of a partially covalent bond between the oxygen atoms and the metal surface (charge donation) and a charge backdonation from the metal into the lowest unoccupied molecular orbital (LUMO).[32] Since the charge backdonation brings the LUMO into resonance with the Fermi-energy (see Figure S1), these systems are also known as "Fermi-level pinned".[1,5,6] At the same time, they exhibit diverse structures with different packing densities on the Ag(111) surface.[26] For clarity, this discussion will only consider the medium-sized 2,7-anthraquinone (labeled A2O further on). The results for the smaller 1,4-benzoquinone and the larger 3,10-pentacenequinone are given in the Supporting Information.

<u>Large geometric reorientation</u>: As a third class, we investigate the situation for tetracyanoethylene (TCNE) on Cu(111). This system is also Fermi-level pinned, but can adsorb in two different states (either flat-lying or upright-standing),[25] which have also been observed experimentally.[33] As we will later, this re-orientation has a quite profound impact on the relation between polymorphism and the adsorption-induced change of the interface work function, $\Delta\phi$.

In the following we will investigate two different deposition conditions. First, we will evaluate the possible range of $\Delta\phi$ that might occur for kinetically trapped phases. After that, we will look at the ranges of $\Delta\phi$ that are accessible if the interface is grown in thermodynamic equilibrium at different temperatures. While in the case of kinetic trapping any (meta)stable polymorph might form, regardless of its formation energy, in thermodynamic equilibrium only the energetically most favorable polymorphs might contribute to the distribution of $\Delta\phi$.

### 3.1. Kinetic Trapping

As a first step, it is useful to explore the range of adsorption-induced work function changes $\Delta\Phi$ that a given interface may exhibit when assuming different structures. By mapping out the energies and $\Delta\Phi$s of the millions of polymorph candidates (which all consist of multiple non-equivalent molecules in unit cells of various shapes and sizes) within our method we can estimate which values of $\Delta\Phi$ can, in principle, occur. The results are shown in Figure 2. All the polymorphs indicated in Figure 2 are possible candidates for kinetically trapped phases, thus the overall possible spread of $\Delta\Phi$ that might be observed for kinetically trapped phases is determined by its total variability (i.e., the width of the distributions in Figure 2).

Interestingly, for none of the systems we observe a strong correlation between $\Delta\Phi$ and the adsorption energy per molecule. This indicates that, irrespective of whether the system shows a large charge transfer or not, dipole-dipole interactions do not play a major role for the formation energy. Nonetheless, the three prototypical systems show clearly different behavior, which is directly related to the molecular orientations and the nature of the interactions with the substrate. For the flat-lying, Fermi level pinned system A2O (shown in green in Figure 2), we find that $\Delta\Phi$ hardly depends on the polymorph at all. This holds independent of the molecular coverage (within certain limitations, compare also Figure S6). Between all polymorph candidates we sampled and predicted, $\Delta\Phi$ only varies by as little as 250 meV. Qualitatively, we observe the same behavior for all flat-lying, Fermi-level pinned systems, i.e., also for TCNE/Cu (orange in Figure 2), as well as for B2O/Ag and P2O/Ag (both in the Supporting Section 2). This is a natural consequence of the interaction mechanism between substrate and adsorbate, where collective electrostatic effects lead to the formation of an interface dipole that is just large enough to shift the LUMO of the molecule into resonance with the Fermi level.[34] For flat-lying molecules, the LUMO is essentially independent of the structure, leading to an almost constant ΔΦ. We tentatively attribute the small remaining variability of ΔΦ to minute changes of the LUMO energy due to intermolecular interactions and wave-function overlap.

For naphthalene on Cu(111), no such compensating effect exists. Naphthalene induces its ΔΦ via Pauli-pushback, i.e., the displacement of substrate electron density, which is different for each adsorption site. As the same electron density cannot be displaced twice, the dipole induced by each molecule is also affected by the presence of other molecules close by, which is reminiscent of (but physically different from) depolarization. As it turns out, naphthalene on Cu can form multiple structures with very similar energies, but notably different ΔΦ. In our data, the variability of ΔΦ spans up to 400 meV.

The upright-standing structures of TCNE are also Fermi-level pinned (see Figure 1d). However, they show two striking differences to their flat-lying counterparts. First, ΔΦ is substantially larger (by ~3 eV, shown in red in Figure 2). Second, between the different polymorphs, the variability is extremely high, spanning as much as 1.7 eV. Like for naphthalene, also here we find multiple structures that have the same energy per molecule, but substantially different ΔΦ. The reason for this effect is that the CN-groups of this molecule, in the upright-standing geometry, create a substantial surface dipole for the organic layer.[35] This (molecular) surface dipole substantially shifts the LUMO relative to the substrate, which must then be compensated by a larger charge-transfer-induced interface dipole.[36] The resulting ΔΦ is influenced by the density of those surface dipoles. At the same time there are many polymorph candidates with different coverages, but similar adsorption energies (see Figure 3d top) leading to a large ΔΦ spread even for beneficial energies.

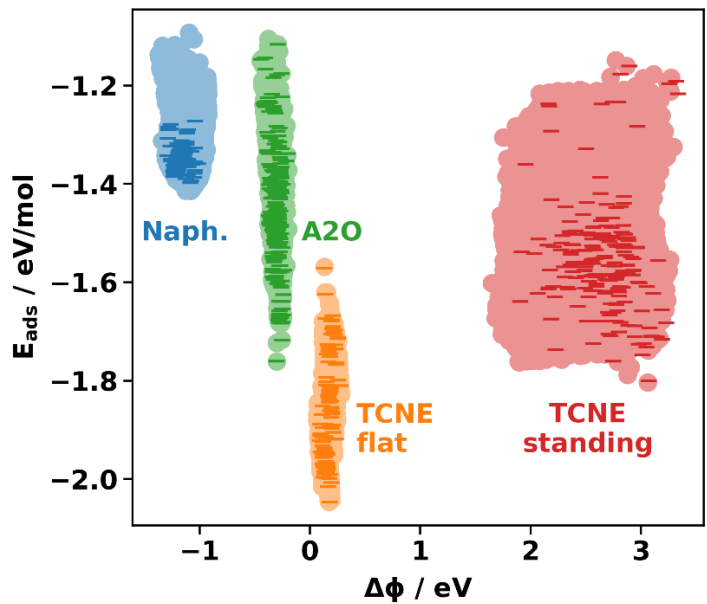

**Figure 2: Adsorption energy per molecule for all polymorph candidates used throughout this study versus their adsorption-induced work function change Δϕ. The horizontal lines indicate the DFT calculations performed to predict all the polymorph candidates. Each colored dot (most of which are overlapping) indicates one of those candidates.**

We note that the energetically favorable polymorphs, of course, have different coverages. The top row of Figure 3 visualizes the adsorption energies in relation to the areas per molecule. Due to purely repulsive intermolecular interactions[16], naphthalene exhibits its energetically most favorable polymorphs at the lowest densities used in this study. For the other systems, attractive intermolecular interactions [25,26] lead to energetically favorable polymorphs within the considered area ranges. The area dependence plays only a minor role for kinetic trapping as polymorphs of any density might be trapped, but it will become important for the thermal occupation.

In short, we find that for flat-lying, Fermi-level pinned systems ΔΦ is hardly dependent on the polymorph. A more notable variation is observed for physisorbed systems. The by far largest dependence of ΔΦ is observed for the system which contains a surface dipole, and where kinetic trapping may readily lead to polymorphs with different densities.

### 3.2. Thermal Occupation

As a next step, we explore in the most probable ΔΦ for interfaces in thermodynamic equilibrium as well as the "smearing" of the work function at elevated temperatures (i.e., ΔΦ distribution due to the thermal occupation of multiple, energetically favorable, polymorphs with different work functions). When the deposition approaches full monolayer coverage, the molecules will try to optimize the Gibbs free energy of adsorption γ.

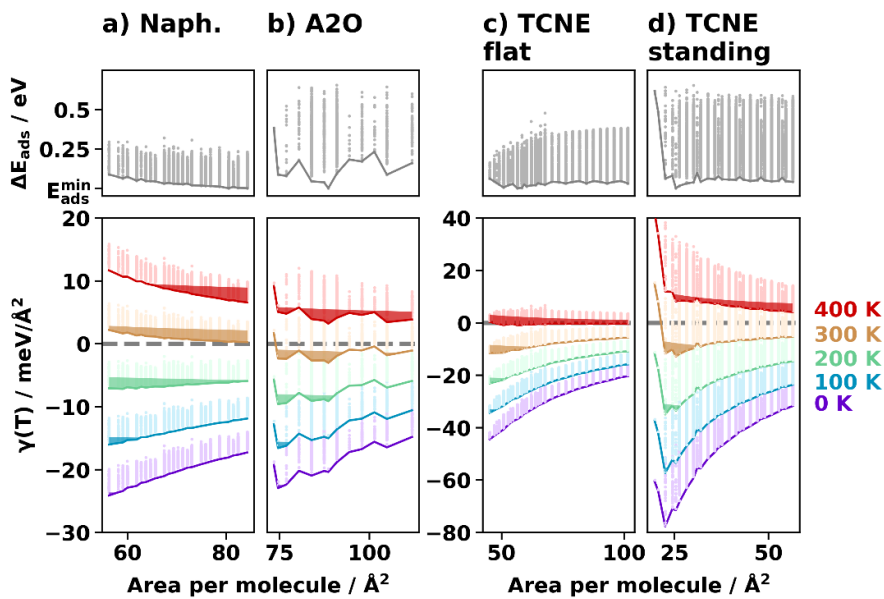

**Figure 3:** Visualization of the adsorption energy relative to the energetically best polymorph candidate for each system $\Delta E_{ads}$ (upper row) and the Gibbs free energy of adsorption γ for all investigated systems as function of their packing density (area per molecule) at different temperatures T (bottom row). The colored line is a guide to the eye, connecting the energetically most favorable polymorphs at each area. The shaded areas indicate areas with a probability higher than 1/100 relative to the Boltzmann probability of the best polymorph at that temperature and a pressure of $10^{-9}$ mbar.

Figure 3 shows the influence of temperature on the Gibbs free energy of adsorption for all polymorph candidates. Generally, at low temperatures, the energetically best polymorphs are located at small areas, i.e. densely packed. Even naphthalene, with its purely repulsive interactions, shows a dense packing at monolayer coverage. With increasing temperature, polymorphs with looser packing (larger area) become energetically more stable compared to their dense competitors. At the same time, the increased temperature leads to more different polymorphs becoming thermally acessible (shaded areas in Figure 3).

Consequently, for naphthalene looser packed structures become the energetically most favorable. For A2O, the transition to loose packed structures happens at higher temperatures, which are often beyond the desorption limit ($\gamma = 0$), indicated in Figure 3 with a dashed line. For TCNE, at low temperatures the standing polymorphs are energetically more favorable than any lying polymorph (being also more densely packed). When increasing the temperature to approximately 250 K, flat polymorphs start to become the energetically more favorable ones, causing a phase transition from standing to lying.

Having the Gibbs free energy of adsorption as an energetic measure to quantify the polymorphs' stabilities at increased temperatures, we can ask the question: How will the expected $\Delta\Phi$ change with changing temperature? For this question we utilize the following theoretical experiment: A macroscopic metal sample is placed inside an ultra high vacuum chamber with a constant pressure (here we use $10^{-9}$ mbar ($10^{-7}$ Pa)). We deposit adsorbate molecules at a low temperature such that a full monolayer is formed and a gas phase reservoir of molecules forms. In this experiment, at zero Kelvin the chemical potential of the molecules in the gas phase, µ, can be neglected, thus only the polymorph which is energetically most favorable in terms of energy per area will form (if there are multiple energetically degenerate polymorphs, there might be an equivalent mixture of all of them). Increasing the temperature in this experiment will (i) make looser packed structures energetically more favorable due to an increasing contribution of µ to the Gibbs free energy of adsorption and (ii) allow more and more energetically less favorable polymorphs to be occupied due to thermal occupation. Details of the thermal occupation estimation are presented in Supporting Section 5.

Figure 4 visualizes the evolution of $\Delta\Phi$ when changing the temperature in this hypothetical experiment. For the visualization, we calculated the Boltzmann weights for each polymorph at each temperature. To account for theoretical uncertainties, we modelled the $\Delta\phi$ of each polymorph as a gaussian distribution, centered at the polymorph's $\Delta\phi$ with a standard deviation of 10 meV (which roughly corresponds to the prediction uncertainties of our machine-learning approach). The $\Delta\phi$ distribution (indicated with shaded curves) is a sum of all those polymorph contributions weighted by their corresponding Boltzmann weights. For all systems, the empty surface is included as a separate polymorph with an effective $\gamma$ of zero. To increase the visibility of low-weight polymorph candidates we also show the distribution on a logarithmic scale (dashed line). The mean and standard deviation of this probability distribution are then indicated via horizontal error bars.

For naphtalene, the work function slightly drifts towards higher values (i.e. lower $\Delta\phi$) for higher temperatures. To explain this, let us briefly reconsider the cause of $\Delta\phi$ when depositing naphthalene on the Cu surface. When these molecules adsorb on the surface, Pauli pushback leads to an effective dipole for each molecule. A change in molecular density on the surface will therefore lead to a reduction of this dipole density and, connected to that, a reduction of $\Delta\phi$. Figure 3a shows that increasing the temperature for naphthalene leads to an increase in the mean area of the best polymorphs which explains the slight drift towards smaller $\Delta\Phi$ at higher temperatures.

For A2O, with increasing temperature more and more polymorph candidates contribute to the mean $\Delta\phi$, but due to the low variability in $\Delta\Phi$ the uncertainty is small, up to the point of desorption where a mixture with the free-surface state leads to a stronger change in work function.

For TCNE we here combine the polymorphs of the lying and the standing systems to get a full picture of the work function evolution. At low temperatures, the standing phases have a more negative Gibbs free energy of adsorption and therefore dominate the work function. Interestingly, despite the fact that there are many energetically similar polymorphs with dissimilar $\Delta\Phi$s (see Figure 2d), here the variability of $\Delta\phi$ within the standing polymorph encompasses only a few 100 meV and is by no means comparable to the one shown in Figure 2. This is because the phase transition to the lying molecules (which itself causes an abrupt, large change in $\Delta\Phi$) happens at around 250K, which is before standing polymorphs with lower density (and corrrespondingly lower dipole density and $\Delta\phi$) could be occupied.The stability of the lying polymorphs is in a temperature range where a large variability in unit cell areas would be possible (compare Figure 3c) which could in principle result in a larger change of $\Delta\phi$. However, like for A2O, the pinning of the molecular LUMO to the metal's Fermi level and the absence of (partial) dipole moments above the pinned level prevents any large change.

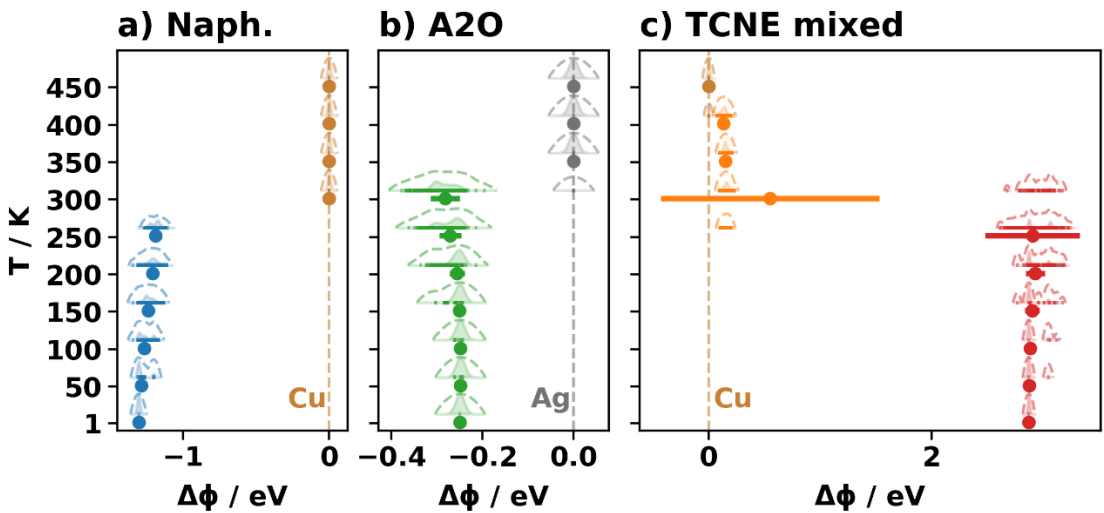

**Figure 4: Expected $\Delta\phi$ for all investigated systems at different temperatures for a molecular background pressure of $10^{-9}$ mbar. The horizontal error bars show the expected work functions and uncertainties at specific temperatures. The shadowed curves indicate the probability distributions obtained by representing each contributing polymorph (small vertical lines) via a gaussian distribution and adding up their contributions weighted by the corresponding Boltzmann weights. The standard deviations of the Gaussians were chosen to be 10 meV. To increase the visibility of low probabilities, the distributions are also plotted on a logarithmic scale (dashed lines). The free-area polymorphs are indicated with the corresponding substrate color and a dashed vertical line at zero $\Delta\phi$.**

## 4. Conclusion

To summarize, we investigated the influence of on-surface polymorphism on the adsorption-induced work function changes on the example of three prototypical metal/organic interfaces: The weakly interacting naphthalene on Cu(111), anthraquinone, which undergoes a charge transfer reaction on the Ag(111) surface, and TCNE, which shows a phase transformation from lying to standing polymorphs upon adsorption on Cu(111). By predicting the work functions of millions of polymorph candidates using DFT and machine learning we gained an overview over the possible work function changes that might be realized by the different systems when kinetically trapped or grown in thermodynamic equilibrium at different temperatures.

When the deposition happens fast, i.e., when any of the polymorph candidates are stabilized via kinetic trapping, the whole variability in $\Delta\Phi$ might be observed experimentally. We found that for flat-lying, Fermi-level pinned interfaces this variability in $\Delta\Phi$ is only small, independent of the surface polymorph that forms. Conversely, for (sub)monolayer systems that are not Fermi-level pinned, or for systems with a large surface dipole (such as upright-standing molecules), polymorphs with very similar adsorption energies show strongly varying $\Delta\Phi$.

If the system is kept in thermodynamic equilibrium, the temperature determines the number of thermally occupied polymorphs and the preferred coverage. Hereby only a large structural change, i.e., the standing up and rehybridization of TCNE, could massively influence the work function (around 3 eV). Within the subclasses, we observed a temperature dependent drift of the work function for the system that was not Fermi-level pinned (naphthalene). But overall, the work function changes observed by thermal effects were only a fraction of the full variability obtainable via kinetic trapping, independent of the interaction mechanisms.

## ASSOCIATED CONTENT

**Supporting Information**

The Supporting Information contains results for the two acenequinones not shown in the main manuscript, details of the prediction process for the polymorph properties, alternative visualizations of the interplay between work function, adsorption energy and temperature as well as charge transfer considerations for the different interface systems and details of the thermal occupation estimation.

**Notes**

The authors declare no competing financial interests.

**Acknowledgments**

We acknowledge fruitful discussions with B. Kirchmayr and J. J. Cartus. Funding through the projects of the Austrian Science Fund (FWF): Y1175-N36 is gratefully acknowledged. Computational results have been achieved using the Vienna Scientific Cluster (VSC).

# References

[1] R. Otero, A.L. Vázquez de Parga, J.M. Gallego, Electronic, structural and chemical effects of charge-transfer at organic/inorganic interfaces, Surf. Sci. Rep. 72 (2017) 105–145. https://doi.org/10.1016/j.surfrep.2017.03.001.

[2] N. Koch, Organic Electronic Devices and Their Functional Interfaces, ChemPhysChem. 8 (2007) 1438–1455. https://doi.org/10.1002/cphc.200700177.

[3] H. Ishii, K. Sugiyama, E. Ito, K. Seki, Energy Level Alignment and Interfacial Electronic Structures at Organic/Metal and Organic/Organic Interfaces, Adv. Mater. 11 (1999) 605–625. https://doi.org/10.1002/(SICI)1521-4095(199906)11:8<605::AID-ADMA605>3.0.CO;2-Q.

[4] N. Koch, Organic Electronic Devices and Their Functional Interfaces, ChemPhysChem. 8 (2007) 1438–1455. https://doi.org/10.1002/cphc.200700177.

[5] S. Braun, W.R. Salaneck, M. Fahlman, Energy-Level Alignment at Organic/Metal and Organic/Organic Interfaces, Adv. Mater. 21 (2009) 1450–1472. https://doi.org/10.1002/adma.200802893.

[6] E. Zojer, T.C. Taucher, O.T. Hofmann, The Impact of Dipolar Layers on the Electronic Properties of Organic/Inorganic Hybrid Interfaces, Adv. Mater. Interfaces. 6 (2019) 1900581. https://doi.org/10.1002/admi.201900581.

[7] D. Käfer, L. Ruppel, G. Witte, Growth of pentacene on clean and modified gold surfaces, Phys. Rev. B - Condens. Matter Mater. Phys. 75 (2007) 085309. https://doi.org/10.1103/PhysRevB.75.085309.

[8] N. Koch, A. Gerlach, S. Duhm, H. Glowatzki, G. Heimel, A. Vollmer, Y. Sakamoto, T. Suzuki, J. Zegenhagen, J.P. Rabe, F. Schreiber, Adsorption-induced intramolecular dipole: Correlating molecular conformation and interface electronic structure, J. Am. Chem. Soc. 130 (2008) 7300–7304. https://doi.org/10.1021/ja800286k.

[9] W.H. Lee, J. Park, S.H. Sim, S. Lim, K.S. Kim, B.H. Hong, K. Cho, Surface-directed molecular assembly of pentacene on monolayer graphene for high-performance organic transistors, J. Am. Chem. Soc. 133 (2011) 4447–4454. https://doi.org/10.1021/ja1097463.

[10] G. Bavdek, A. Cossaro, D. Cvetko, C. Africh, C. Blasetti, F. Esch, A. Morgante, L. Floreano, Pentacene nanorails on Au(110), Langmuir. 24 (2008) 767–772. https://doi.org/10.1021/la702004z.

[11] A.O.F. Jones, B. Chattopadhyay, Y.H. Geerts, R. Resel, Substrate-Induced and Thin-Film Phases: Polymorphism of Organic Materials on Surfaces, Adv. Funct. Mater. 26 (2016) 2233–2255. https://doi.org/10.1002/adfm.201503169.

[12] Y. Zou, L. Kilian, A. Schöll, T. Schmidt, R. Fink, E. Umbach, Chemical bonding of PTCDA on Ag surfaces and the formation of interface states, Surf. Sci. 600 (2006) 1240–1251. https://doi.org/10.1016/j.susc.2005.12.050.

[13] S. Duhm, A. Gerlach, I. Salzmann, B. Bröker, R.L. Johnson, F. Schreiber, N. Koch, PTCDA on Au(1 1 1), Ag(1 1 1) and Cu(1 1 1): Correlation of interface charge transfer to bonding distance, Org. Electron. 9 (2008) 111–118. https://doi.org/10.1016/j.orgel.2007.10.004.

[14] A.O.F. Jones, B. Chattopadhyay, Y.H. Geerts, R. Resel, Substrate-induced and thin-film phases: Polymorphism of organic materials on surfaces, Adv. Funct. Mater. 26 (2016) 2233–2255. https://doi.org/10.1002/adfm.201503169.

[15] J. Nyman, G.M. Day, Static and lattice vibrational energy differences between polymorphs, CrystEngComm. 17 (2015) 5154–5165. https://doi.org/10.1039/c5ce00045a.

[16] L. Hörmann, A. Jeindl, A.T. Egger, M. Scherbela, O.T. Hofmann, SAMPLE: Surface structure search enabled by coarse graining and statistical learning, Comput. Phys. Commun. 244 (2019) 143–155. https://doi.org/10.1016/J.CPC.2019.06.010.

[17] A. Jeindl, L. Hörmann, O.T. Hofmann, DFT calculations for the Structure Search of para-benzoquinone on Ag(111), (2021). https://doi.org/10.17172/NOMAD/2021.03.09-3.

[18] A. Jeindl, L. Hörmann, O.T. Hofmann, DFT calculations for the Structure Search of anthracenequinone on Ag(111), (2021). https://doi.org/10.17172/NOMAD/2021.03.09-2,.

[19] A. Jeindl, L. Hörmann, O.T. Hofmann, DFT calculations for the Structure Search of pentacenequinone on Ag(111), (2021). https://doi.org/10.17172/NOMAD/2021.03.09-1.

[20] L. Hörmann, A. Jeindl, O.T. Hofmann, DFT Calculations for the Structure Search of naphthalene on Cu(111), (2021).

https://doi.org/10.17172/NOMAD/2021.06.30-1.

[21] A. Jeindl, L. Hörmann, O.T. Hofmann, DFT calculations for the Structure Search of tetracyanoethylene on Cu(111), (2021). https://doi.org/10.17172/NOMAD/2021.06.30-2.

[22] J.P. Perdew, K. Burke, M. Ernzerhof, Generalized Gradient Approximation Made Simple, Phys. Rev. Lett. 77 (1996) 3865–3868. https://doi.org/10.1103/PhysRevLett.77.3865.

[23] A. Tkatchenko, M. Scheffler, Accurate Molecular Van Der Waals Interactions from Ground-State Electron Density and Free-Atom Reference Data, Phys. Rev. Lett. 102 (2009) 073005. https://doi.org/10.1103/PhysRevLett.102.073005.

[24] V.G. Ruiz, W. Liu, E. Zojer, M. Scheffler, A. Tkatchenko, Density-Functional Theory with Screened van der Waals Interactions for the Modeling of Hybrid Inorganic-Organic Systems, Phys. Rev. Lett. 108 (2012) 146103. https://doi.org/10.1103/PhysRevLett.108.146103.

[25] A.T. Egger, L. Hörmann, A. Jeindl, M. Scherbela, V. Obersteiner, M. Todorović, P. Rinke, O.T. Hofmann, Charge Transfer into Organic Thin Films: A Deeper Insight through Machine-Learning-Assisted Structure Search, Adv. Sci. 7 (2020) 2000992. https://doi.org/10.1002/advs.202000992.

[26] A. Jeindl, J. Domke, L. Hörmann, F. Sojka, R. Forker, T. Fritz, O.T. Hofmann, Nonintuitive Surface Self-Assembly of Functionalized Molecules on Ag(111), ACS Nano. (2021). https://doi.org/10.1021/acsnano.0c10065.

[27] L. Hörmann, A. Jeindl, A.T. Egger, M. Scherbela, O.T. Hofmann, SAMPLE: Surface structure search enabled by coarse graining and statistical learning, Comput. Phys. Commun. 244 (2019) 143–155. https://doi.org/10.1016/J.CPC.2019.06.010.

[28] K. Reuter, M. Scheffler, Composition, structure, and stability of RuO2(110) as a function of oxygen pressure, Phys. Rev. B - Condens. Matter Mater. Phys. 65 (2002) 1–11. https://doi.org/10.1103/PhysRevB.65.035406.

[29] P. Herrmann, G. Heimel, Structure and Stoichiometry Prediction of Surfaces Reacting with Multicomponent Gases, Adv. Mater. 27 (2015) 255–260. https://doi.org/10.1002/adma.201404187.

[30] T. Yamada, M. Shibuta, Y. Ami, Y. Takano, A. Nonaka, K. Miyakubo, T. Munakata, Novel growth of naphthalene overlayer on Cu(111) studied by STM, LEED, and 2PPE, J. Phys. Chem. C. 114 (2010) 13334–13339. https://doi.org/10.1021/jp1045194.

[31] R. Forker, J. Peuker, M. Meissner, F. Sojka, T. Ueba, T. Yamada, H.S. Kato, T. Munakata, T. Fritz, The complex polymorphism and thermodynamic behavior of a seemingly simple system: Naphthalene on Cu(111), Langmuir. 30 (2014) 14163–14170. https://doi.org/10.1021/la503146w.

[32] G. Heimel, S. Duhm, I. Salzmann, A. Gerlach, A. Strozecka, J. Niederhausen, C. Bürker, T. Hosokai, I. Fernandez-Torrente, G. Schulze, S. Winkler, A. Wilke, R. Schlesinger, J. Frisch, B. Bröker, A. Vollmer, B. Detlefs, J. Pflaum, S. Kera, K.J. Franke, N. Ueno, J.I. Pascual, F. Schreiber, N. Koch, Charged and metallic molecular monolayers through surface-induced aromatic stabilization, Nat. Chem. 5 (2013) 187–194. https://doi.org/10.1038/nchem.1572.

[33] W. Erley, Reflection-absorption infrared spectroscopy of tetracyanoethylene adsorbed on copper(111): observation of vibronic interaction, J. Phys. Chem. 91 (1987) 6092–6094. https://doi.org/10.1021/j100308a007.

[34] O.T. Hofmann, V. Atalla, N. Moll, P. Rinke, M. Scheffler, Interface dipoles of organic molecules on Ag(111) in hybrid density-functional theory, New J. Phys. 15 (2013) 123028. https://doi.org/10.1088/1367-2630/15/12/123028.

[35] S. Duhm, G. Heimel, I. Salzmann, H. Glowatzki, R.L. Johnson, A. Vollmer, J.P. Rabe, N. Koch, Orientation-dependent ionization energies and interface dipoles in ordered molecular assemblies, Nat. Mater. 7 (2008) 326–332. https://doi.org/10.1038/nmat2119.

[36] O.T. Hofmann, D.A. Egger, E. Zojer, Work-function modification beyond pinning: When do molecular dipoles count?, Nano Lett. 10 (2010) 4369–4374. https://doi.org/10.1021/nl101874k.

# Supporting Information to "How much does surface polymorphism influence the work function of organic/metal interfaces?"

Andreas Jeindl[1], Lukas Hörmann[1] and Oliver T. Hofmann[1]*

[1] Institute of Solid State Physics, NAWI Graz, Graz University of Technology, Petersgasse 16, 8010 Graz, Austria

**Supporting Section 1: Fermi Level Pinning of the Investigated Systems**

In addition to Figure S1 provides the Molecular Orbital Density of States (MODOS) for the energetically best geometric building block (local geometry) on the respective metal surface. It shows that all systems, except Naphthalene on Cu(111) are Fermi-level pinned. The pinning for P2O occurs not only to the LUMO but also the LUMO+1 due to coincidental energetic similarity between those two orbitals.

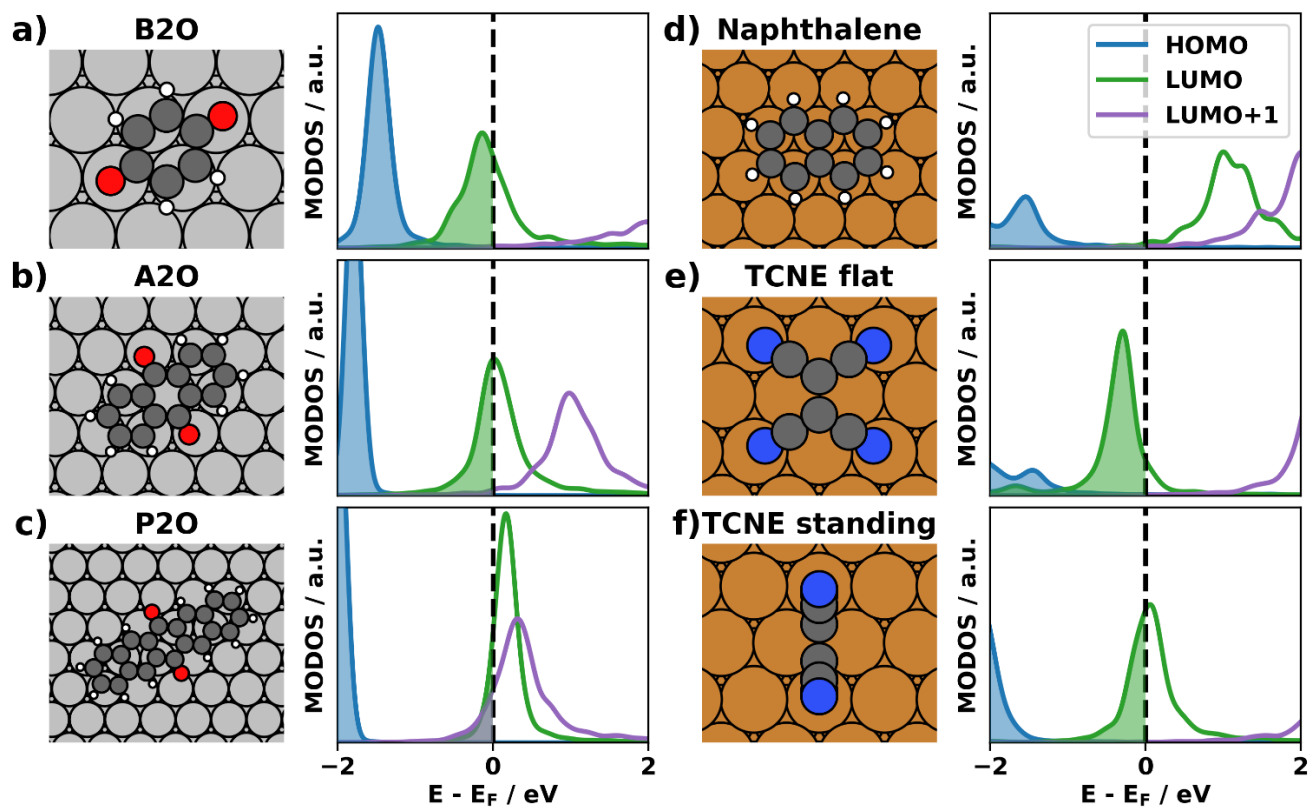

**Figure S1:** Geometric structure and molecular orbital density of states for. a) para-benzoquinone (B2O) on Ag(111), b) 2,7-anthraquinone (A2O) on Ag(111) c) 3,10-pentacenequinone (P2O) on Ag(111), d) Naphthalene on Ag(111), e-f) tetracyanoethylene (TCNE) on Cu(111). A partly filled LUMO (and, in the case of P2O, LUMO+1) indicates Fermi level pinning of the system.

**Supporting Section 2: Evaluation of Additional Acenequinones on Ag(111)**

In addition to the already discussed anthraquinone, here we perform the same analyses for the smaller molecule benzoquinone and the larger molecule pentacenequinone. Figures S2 to S4 contain the same analyses as their counterparts in the main manuscript, but for different systems. Throughout the rest of this supporting information, we will provide the details for all six systems.

Figure S2 indicates that the variability of Δϕ for P2O is in the same range as for A2O (~250 meV). For B2O, it is slightly larger (~550 meV). However, this is likely an artifact of too closely packed polymorph candidates at higher energies.

Figure S3 shows that for B2O the best polymorph candidates are at the densely packed end of the coverage spectrum all the way up until desorption. For P2O, a transition from densely packed to loosely packed structures occurs before the molecules desorb (above γ=0). Due to the Fermi level pinning of the systems this does, however, not influence the mean work function (and its spread) at elevated temperatures (Figure S4)

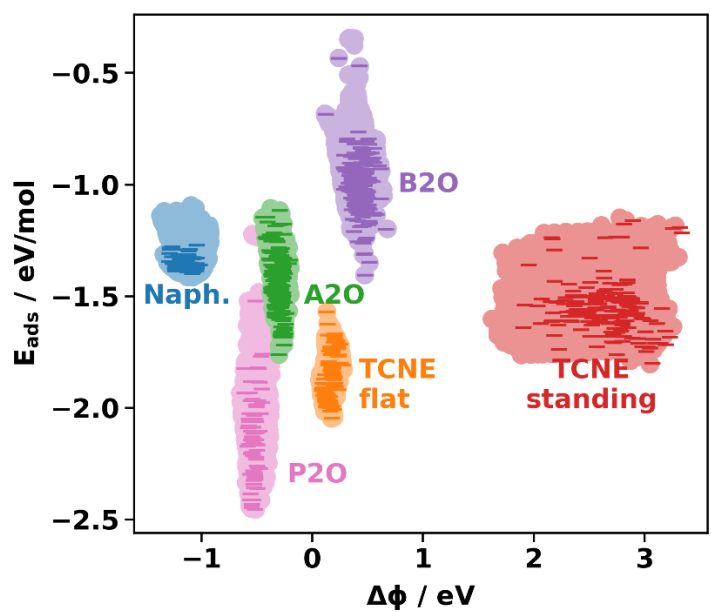

**Figure S2: Adsorption energy per molecule for all polymorph candidates of all systems relative to their interface work function change Δϕ.**

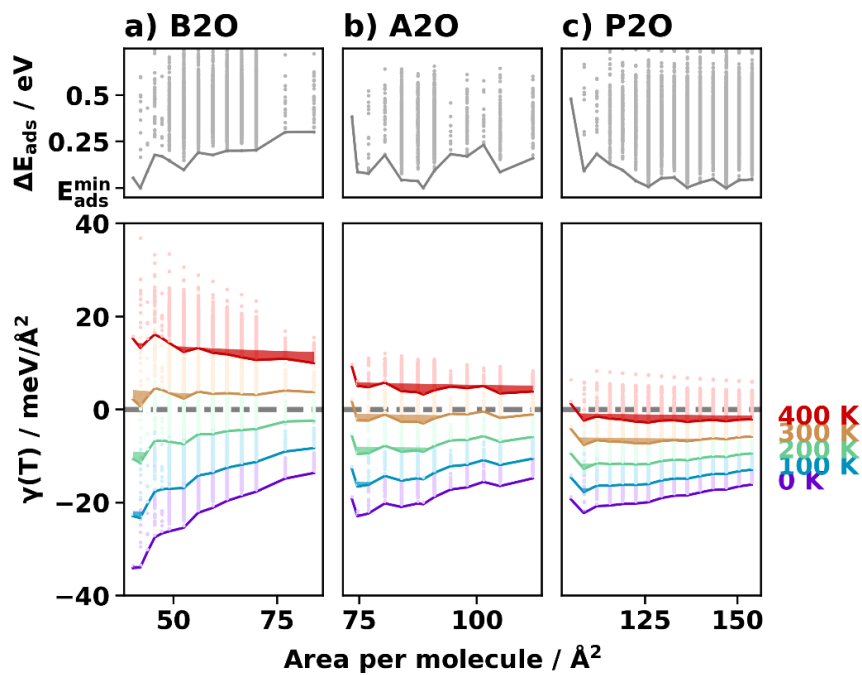

**Figure S3:** Visualization of the adsorption energy relative to the energetically best polymorph candidate for each system $\Delta E_{ads}$ (upper row) and the Gibbs free energy of adsorption $\gamma$ for all investigated systems as function of their packing density (area per molecule) at different temperatures T (lower row). The colored line is a guide to the eye, connecting the energetically most favorable polymorphs at each area. The shaded areas indicate areas with a probability higher than 1/100 relative to the Boltzmann probability of the best polymorph at that temperature and a pressure of $10^{-9}$ mbar.

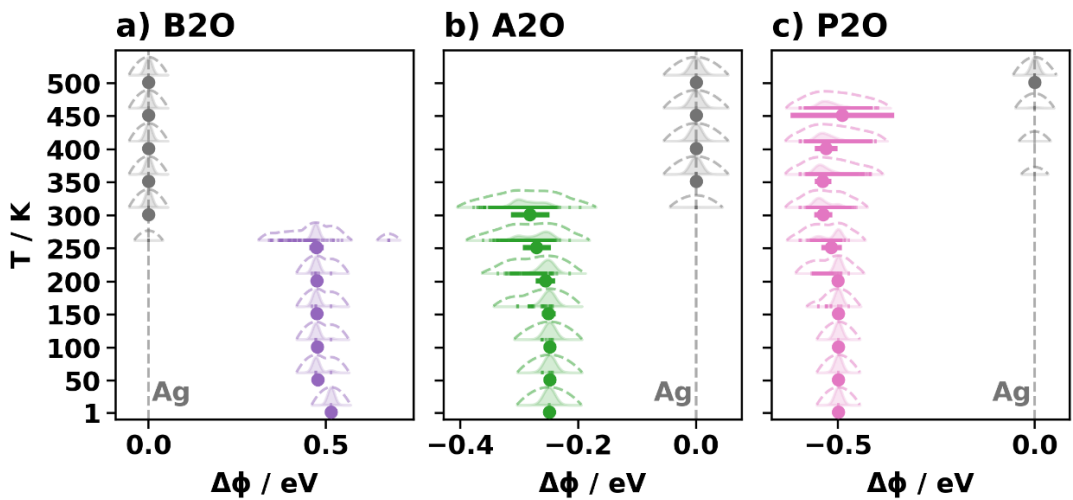

**Figure S4:** Expected $\Delta\phi$ for the three acenequinones systems at different temperatures for a molecular background pressure of $10^{-9}$ mbar. The horizontal error bars show the expected work functions and uncertainties at specific temperatures. The shadowed curves indicate the probability distributions obtained by representing each contributing polymorph (small vertical lines) via a gaussian distribution and adding up their contributions weighted by the corresponding Boltzmann weights. The standard deviations of the Gaussians were chosen to be 10 meV. To increase the visibility of low probabilities, the distributions are also plotted on a logarithmic scale (dashed lines). The free-area polymorphs are indicated with the corresponding substrate color and a dashed vertical line at zero $\Delta\phi$.

## Supporting Section 3: Hyperparameters Used for the SAMPLE Approach

For the SAMPLE approach[1] several hyperparameters are necessary. All of those hyperparameters were varied systematically to maximize the log-likelihood in the Bayesian linear regression formalism. Table S1 contains all optimized hyperparameters used for the prediction of the work function change. The hyper parameters for the prediction of the adsorption energies are the same as for the original publications and are here listed in Table S2. For TCNE and Naphthalene a single atomic species determines the minimum distances between the molecules interacting on the surface, thus a single minimum distance threshold is sufficient. For the acenequinones this is not the case and so multiple minimum distances between different species need to be defined.

Table S1: Hyperparameters of the SAMPLE approach used to predict the interface dipoles

| Hyperparameter | B2O | A2O | P2O | TCNE flat | TCNE standing | Naphthalene |
| --- | --- | --- | --- | --- | --- | --- |
| Single body uncertainty | 0.5 D | 0.5 D | 0.5 D | 0.3 D | 0.2 D | 0.5 D |
| Two-body uncertainty | 0.2 D | 0.2 D | 0.2 D | 0.1 D | 0.1 D | 0.05 D |
| DFT data uncertainty | 0.01 D | 0.01 D | 0.01 D | 0.01 D | 0.05 D | 0.01 D |
| Decay length | 6 Å | 5 Å | 10 Å | 2 Å | 15 Å | 15 Å |
| Decay power | 2 | 2 | 2 | 3 | 2 | 1 |
| Decay length feature space | 12 | 9 | 5 | 12 | 18 | 2 |
| Feature threshold | 0.01 | 0.01 | 0.01 | 0.02 | 0.01 | 0.001 |
| Minimal distance threshold | O↔H: 1.6 Å; O↔O: 2.4 Å, H↔H: 1.6 Å C↔H: 2.3 Å, C↔O: 2.5 Å | | | | 2.6 Å | 1.6 Å |

Table S2: Hyperparameters of the SAMPLE approach used to predict the adsorption energies

| Hyperparameter | B2O | A2O | P2O | TCNE flat | TCNE standing | Naphthalene |
| --- | --- | --- | --- | --- | --- | --- |
| Single body uncertainty | 0.1 eV | 0.1 eV | 0.1 eV | 0.5 eV | 0.5 eV | 0.1 eV |
| Two-body uncertainty | 0.3 eV | 0.3 eV | 0.3 eV | 0.2 eV | 0.2 eV | 0.1 eV |
| DFT data uncertainty | 0.005 eV | 0.005 eV | 0.005 eV | 0.01 eV | 0.01 eV | 0.005 eV |
| Decay length | 5 Å | 5 Å | 10 Å | 1 Å | 2 Å | 5 Å |
| Decay power | 3 | 3 | 3 | 3 | 3 | 2 |
| Decay length feature space | 12 | 9 | 5 | 12 | 1 | 10 |
| Feature threshold | 0.075 | 0.075 | 0.01 | 0.02 | 0.02 | 0.01 |
| Minimal distance threshold | O↔H: 1.6 Å; O↔O: 2.4 Å, H↔H: 1.6 Å C↔H: 2.3 Å, C↔O: 2.5 Å | | | | 2.6 Å | 1.6 Å |

**Supporting Section 4: Temperature Dependence of the Work Function for All Polymorphs**

Figure 2 visualizes the adsorption energy per molecule with respect to the work function. Figure 3 shows the Gibbs free energy in dependence of the area per molecule and temperature. In Figure S5 and S6 we employ two additional visualizations to give an even more thorough insight into the polymorphic behavior. Figure S1 visualizes the work function in dependence of $\gamma$ and the temperature for a gas pressure of $10^{-7}$ Pa, indicating the changes in work function distributions at different temperatures.

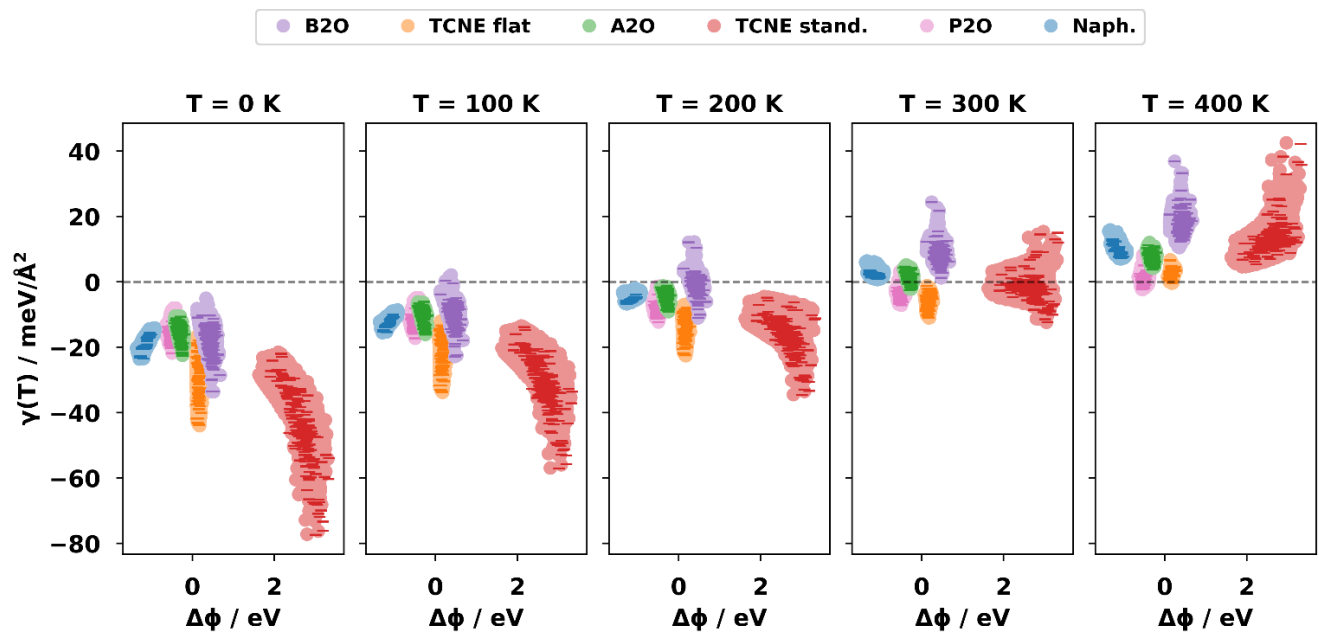

**Figure S5: Visualization of the relation between work function and $\gamma$ at a pressure of $10^{-7}$ Pa and different temperatures**

Figure S6 takes the information of Figure 1, but adds the dependence of the work function and energy on the area, all of that at zero Kelvin. In this figure, the left half of each data point represents $E_{ads}$ and the right half $\gamma(T=0)$.

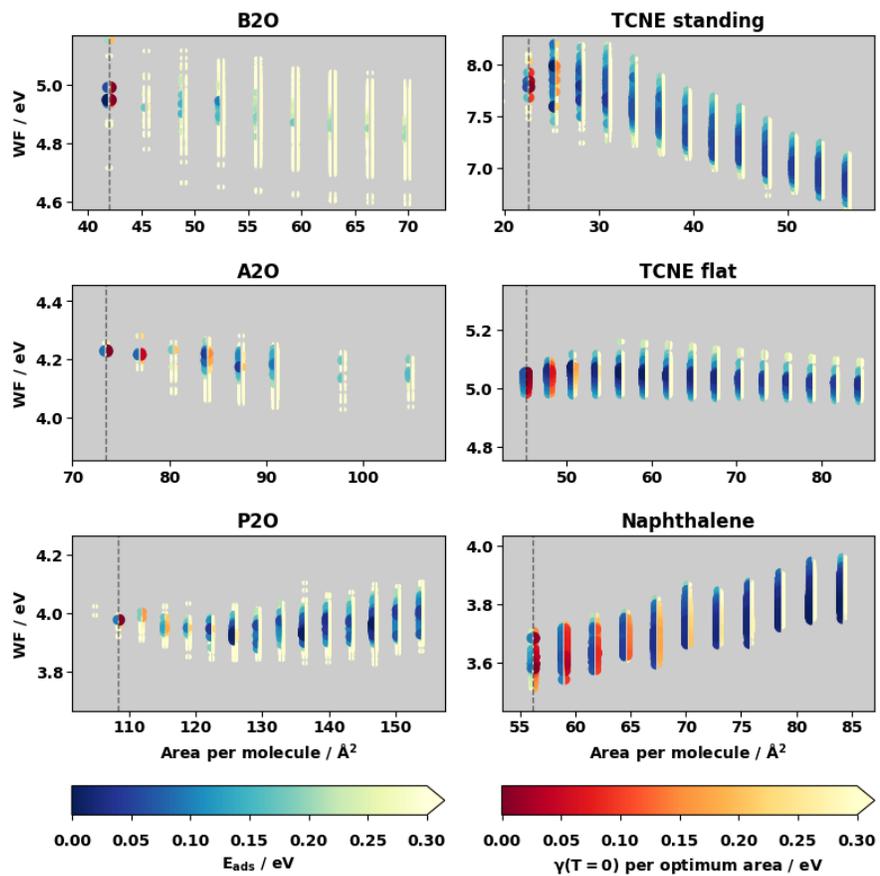

**Figure S6:** Visualization of the work function spread as a function of the coverage. Each dot represents a polymorph candidate. The color of the left half indicates the adsorption energy while the color of the right half indicates the Gibbs free energy at T=0 K

## Supporting Section 5: Calculating the Thermal Occupation

To model thermal occupation, we assume an ensemble where the temperature, pressure and sample area $A_S$ are fixed and the number of molecules adsorbed on the surface N is variable. Within that ensemble, the mean area occupied by a single molecule is given as

$$\bar{A} = \frac{A_S}{N} \quad (S1)$$

The energy gain of a single polymorph occupying the mean area is

$$E_i = \gamma_i * \bar{A} \quad (S2)$$

Additionally, we model the free surface as a "quasi-polymorph" without molecules. This polymorph has an area of $\bar{A}$ and an energy of $\mu$, resulting in a $\gamma$ (and also $E_i$) of zero. At low temperatures, this free-substrate polymorph does not play any role for the effective work function of the system.

The probability of any given polymorph to occupy the mean area is given via its Boltzmann weights

$$p_i = \frac{1}{Z} e^{-\frac{\gamma_i \bar{A}}{k_B T}} \quad (S3)$$

The expectation value of the work function on this hypothetical sample then reads

$$\langle \phi \rangle = \sum_i p_i \phi_i \quad (S4)$$

The mean area implicitly depends on the weights of the individual polymorphs. The number of molecules per mean area for any given polymorph is

$$N_i = \frac{\bar{A}}{A_i} \quad (S5)$$

To update the mean area, we take into account that on average exactly one molecule must be present per mean area

$$\langle N \rangle = \sum_i \frac{\bar{A}}{A_i} p_i \stackrel{m}{=} 1 \quad (S6)$$

which leads to

$$\bar{A} = \frac{1}{\sum \frac{p_i}{A_i}} \quad (S7)$$

As $\bar{A}$ itself depends on the obtained Boltzmann probabilities it can not be calculated a priori. Thus, we optimize it iteratively for every temperature separately via equation (S7).

## References

(1) Hörmann, L.; Jeindl, A.; Egger, A. T.; Scherbela, M.; Hofmann, O. T. SAMPLE: Surface Structure Search Enabled by Coarse Graining and Statistical Learning. *Comput. Phys. Commun.* **2019**, *244*, 143–155. https://doi.org/10.1016/J.CPC.2019.06.010.